\documentclass[aps, superscriptaddress, twocolumn, pre, longbibliography]{revtex4-1}
\usepackage{amsmath,graphicx,color,epsfig,latexsym,bm,ulem,dutchcal,subfigure,float,tikz,eucal, mathpazo,times, braket}
\usepackage[colorlinks=true, linkcolor = blue, urlcolor  = blue, citecolor = blue, anchorcolor = red]{hyperref}   
\newcommand{\abs}[1]{\left\vert#1\right\vert}

\usetikzlibrary{positioning}

\begin{document}
	
	
	\title{Antibunching via cooling by heating}
	\author{M. Tahir Naseem} 
	\author{\"Ozg\"ur E. M\"ustecapl\ifmmode \imath \else \i \fi{}o\ifmmode \breve{g}\else \u{g}\fi{}lu}
	\email{omustecap@ku.edu.tr}
	\affiliation{Department of Physics, Ko\c{c} University, 34450 Sariyer, Istanbul, Turkey}
	
	\begin{abstract}
	We investigate statistics of the photon (phonon) field undergoing linear and nonlinear damping processes. An effective two-photon (phonon) nonlinear ``cooling by heating" process is realized from linear damping by spectral filtering of the heat baths present in the system. 
This cooling process driven by incoherent quantum thermal noise can create quantum states of the photon field. In fact, for high temperatures of the spectrally filtered heat baths, sub-Poissonian statistics with strong antibunching in the photon (phonon) field are reported. 
	 This notion of the emergence and control of quantumness by incoherent thermal quantum noise is applied to a quantum system comprising of a two-level system and a harmonic oscillator or analogous optomechanical setting. Our analysis may provide a promising direction for the preparation and protection of quantum features via nonlinear damping that can be controlled with incoherent thermal quantum noise.
	\end{abstract}
	
	\maketitle

	\section{Introduction}\label{sec:model}
	
The generation and manipulation of the quantum states of light and matter are the quintessence of quantum technologies~\cite{Kurizki3866, BROWNE20172, Moreau2019}. The antibunching phenomenon is a dramatic demonstration of the corpuscular behavior of quantum fields~\cite{PhysRevA.82.013824, PhysRevA.41.475}. The paradigmatic system where antibunching has been proposed~\cite{PhysRevLett.79.1467} and observed is cavity QED~\cite{Birnbaum2005, doi:10.1126/science.1152261, PhysRevLett.118.133604}. Photon antibunching has also been realized in circuit QED~\cite{PhysRevLett.107.053602, PhysRevLett.106.243601} and photonic crystals ~\cite{PhysRevLett.114.233601}. A manifestation of photon antibunching is photon blockade, in which the presence of a single photon in a resonator hinders the transmission of the second one~\cite{PhysRevLett.79.1467}. The physical mechanism behind the photon blockade is nonlinear light-matter interaction making the energy-level spacing anharmonic~\cite{Birnbaum2005}. Other systems, such as circuit QED~\cite{PhysRevA.89.043818}, cavity QED~\cite{PhysRevLett.109.193602}, optomechanical resonators~\cite{PhysRevLett.107.063601, PhysRevA.88.023853, PhysRevA.92.033806, PhysRevLett.121.153601}, and cavity-emitter systems~\cite{PhysRevLett.111.247401, PhysRevA.97.063844, PhysRevA.98.053801, PhysRevLett.122.243602}, have been proposed to realize photon blockade, too. More recently, the possibility of antibunching of phonons in optomechanics has been attracted much attention~\cite{PhysRevA.82.021806, PhysRevA.93.063861, PhysRevA.96.013861, PhysRevA.99.013804}.

Conventional photon blockade occurs when a coherent light enters an optical medium with nonlinearity stronger than the dissipation so that anharmonic level spacing between single and two-photon transitions can be resolved beyond their linewidths; the output of such a system exhibits photon antibunching and can be used as a single-photon source~\cite{RevModPhys.54.1061}. The photon blockade effect was demonstrated in various systems, including cavity QED~\cite{Birnbaum2005}, and circuit QED~\cite{PhysRevLett.107.053602}. To alleviate the requirement of large nonlinearity, an alternative, so-called unconventional photon blockade strategy of using quantum interference of different excitation pathways has been considered~\cite{PhysRevLett.104.183601, PhysRevA.94.013815, PhysRevA.96.053810} in multi-mode driven-dissipative systems. Phonon antibunching has been mainly discussed in optomechanical oscillators~\cite{PhysRevA.95.053844}, hybrid nanomechanical resonator systems~\cite{PhysRevA.82.032101}, and micro/nanoelectromechanical resonators~\cite{Guan_2017}, where thermal fluctuations decrease the antibunching effect, and at high temperatures, phonons become completely bunched. Analogous to photons, both conventional and unconventional routes to phonon blockade have been proposed. Destructive interference of two-phonon excitation pathways can be used for phonon blockade in nanomechanical resonators~\cite{PhysRevA.94.063853} or weakly coupled mechanical oscillators~\cite{Sarma2018}. Still, it is restricted to ultracold temperatures as the phonon blockade is highly fragile with thermal noise.

In this paper, we ask if the quantum interference for unconventional phonon or photon blockade can be realized between the dissipative pathways, and more significantly, if such a mechanism can be more pronounced at higher temperatures. We present positive answers to these questions by a simple filter engineering of the thermal bath spectrum, which can be significant for high-temperature single-photon and phonon sources that can operate solely by thermal noise. Specifically, we propose an unconventional photon/phonon antibunching scheme based on “cooling by heating”~\cite{PhysRevLett.108.120602, Naseem2021} method, where spectral filtering of the thermal baths results in an effective two-photon/phonon damping. Bath spectrum filtering has previously been shown to enhance the performance of certain thermal tasks~\cite{Naseem2021, PhysRevLett.109.090601, PhysRevE.87.012140, PhysRevE.90.022102, Ghosh12156, PhysRevE.99.042121, PhysRevResearch.2.033285, Naseem_2020}. {\color{blue}In addition, various interesting quantum effects induced by mere thermal driving in quantum optical systems have been proposed, such as lasing in quantum heat engines~\cite{PhysRevE.80.061129, Mari_2015, PhysRevE.96.062120, Naseem:19}.}

Our scheme is applicable to both photon and phonon antibunching. We consider two example generic models to show the validity of our scheme. The first model describes a harmonic oscillator coupled to a two-level system; the second model consists of two resonators with optomechanical coupling. These models can be realized with the electromechanical~\cite{PhysRevA.93.063861, LaHaye2009} or circuit QED setups~\cite{PhysRevLett.115.203601, PhysRevB.93.134501}, where the resonator fields correspond to either photons or phonons, respectively. We find high temperatures make the antibunching stronger. Antibunching is a paradigmatic signature of “quantumness“~\cite{scully_zubairy_1997}, whose emergence at higher temperatures is fundamentally significant. In addition, the elimination of degrading effects of thermal noise on single-photon and phonon sources can make our results significant for practical quantum technology applications, too. We explain our counter-intuitive results using quantum interference of dissipative pathways and bath spectrum engineering, making the two-photon decay the dominant mechanism of dissipation.

In addition to the capacity of high-temperature operation, there are other critical differences between our scheme and the previous proposals~\cite{RevModPhys.54.1061} on antibunching induced by two-photon absorption (cooling): (i) In our scheme, antibunching of either photon or phonon field can be realized on the contrary, previous proposals are limited to photon antibunching. (ii) We get effective nonlinear damping from the linear system-bath couplings by employing a bath spectrum filter. (iii) Our scheme is based on cooling by heating~\cite{PhysRevLett.108.120602, Naseem2021}, in which mere incoherent thermal drive produces a cooling effect. Accordingly, the rate of this cooling process can be tuned by the temperatures of the thermal baths. In previous proposals, such control was impossible because the environment needs to be considered at zero temperature to realize cooling. (iv) Our proposal can be realized using different platforms, for example, circuit QED~\cite{PhysRevLett.115.203601, PhysRevLett.120.227702, Bothner2021}, electro-mechanical systems~\cite{LaHaye2009}, and various realizations of optomechanical systems~\cite{RevModPhys.86.1391}.

The rest of the paper is organized as follows: Sec.~\ref{sec:MS} we present the model system, and Sec.~\ref{sec:MA} describes the model analysis. In Sec.~\ref{sec:ME}, and Sec.~\ref{sec:FP} we derive the master equation and Fokker-Plank equation for the system, respectively. The results of two-photon cooling by heating and antibunching are presented in Sec.~\ref{sec:results}. Finally, conclusions of this paper are given in Sec.~\ref{sec:conc}.

\section{Model system}\label{sec:MS}
\begin{figure}[t!]
  \centering
  \includegraphics[scale=0.32]{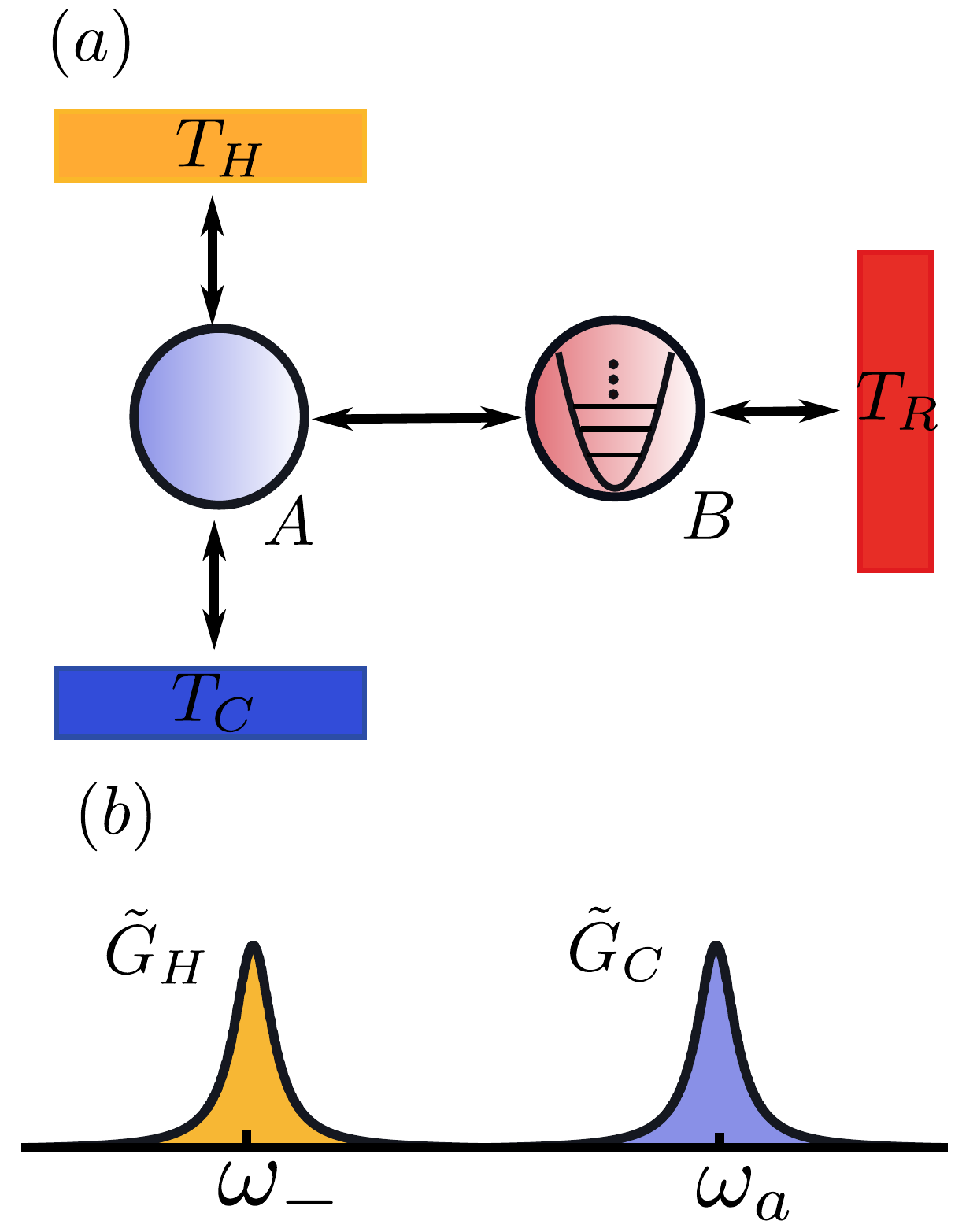}
  \caption{(a) Model description. Our proposal is based on two subsystems A and B interacting via an {\it{energy-field}} (optomechanical-like) interaction. The isolated system $A+B$ may have one of the following forms: (i) subsystem A is a two-level system coupled to a resonator B. For example, a qubit longitudinally coupled to a microwave resonator~\cite{PhysRevLett.115.203601, PhysRevB.93.134501}. (ii) Both subsystems A and B are resonators coupled via an optomechanical-like interaction. For example, an optical cavity mode A coupled to a micromechanical resonator B~\cite{RevModPhys.86.1391}. The subsystems B and A are coupled to one and two thermal baths, respectively. All the baths are independent and can attain any finite non-negative temperature. (b) Spectrally separated baths spectra. $\tilde{G}_{H}$ and $\tilde{G}_{C}$ are the filtered baths spectra of the hot and cold baths, respectively. The center and width of the spectra can be controlled by filter frequency and system-baths coupling rates, respectively. $\omega_{a}$ and $\omega_{b}$ are the frequencies of subsystems A and B, respectively, and $\omega_{-}=\omega_{a}-2\omega_{b}$.}
  \label{fig:fig1}
\end{figure}

We consider a setup consisting of two subsystems interacting via {\it{longitudinal coupling}}, i.e., the energy of subsystem A is coupled to the field of the B. In addition, subsystem A is coupled to two independent thermal baths of temperatures $T_{\alpha=H, C}$ and B is coupled to a single thermal bath of temperature $T_{B}$. The schematic illustration of the model is shown in Fig.~\ref{fig:fig1}(a). The total Hamiltonian of the system can be written as (we take $\hbar=1$)
\begin{equation}
\hat{H}_{T} = \hat{H}_{A+B} + \hat{H}_{E_{j}} + \hat{H}_{A-E_{\alpha}} + \hat{H}_{B-E_{R}}.
\end{equation}
Here, the first term represents the energy of the isolated system $A+B$. The free Hamiltonian of the independent thermal baths is given by the second term
\begin{equation}\label{eq:bathHam}
\hat{H}_{E_{j}} = \sum_{k,j} \omega_{k, j} \hat{c}^{\dagger}_{k,j}\hat{c}_{k,j}.
\end{equation} 
Here, $\hat{c}_{k,j}$ ($\hat{c}^{\dagger}_{k,j}$), and $\omega_{k, j}$ is the annihilation (creation) operator, and frequency of the $k$-th bath mode, respectively. In Eq.~(\ref{eq:bathHam}), the sum is taken over the infinite number of these modes indexed by $k$, and $j= H, C, and$ $R$ represents the thermal baths coupled to subsystem A and B, respectively. The interaction of the isolated system ($S=A+B$) with the baths is given by 
\begin{align}\label{eq:sys-bath}
\hat{H}_{S-E_{j}} &= \sum_{k, \alpha} g_{k, \alpha} (\hat{a} + \hat{a}^{\dagger})(\hat{c}_{k, \alpha} + \hat{c}^{\dagger}_{k, \alpha}) \\ \nonumber   &\quad\qquad +\sum_{k, R} g_{k, R} (\hat{b} + \hat{b}^{\dagger})(\hat{c}_{k, R} + \hat{c}^{\dagger}_{k, R}),
\end{align}
$\hat{a}$ ($\hat{a}^{\dagger}$), and $\hat{b}$ ($\hat{b}^{\dagger}$) being the annihilation (creation) operators of subsystems A and B, respectively.

The Hamiltonian $\hat{H}_{A+B}$ of the isolated system may have one of the following forms: \\
(i) If the energy of a two-level system (A) is coupled to a resonator (B) via its longitudinal degree of freedom~\cite{LaHaye2009, PhysRevLett.115.203601, PhysRevB.93.134501}, then 
\begin{equation}\label{eq:TLS-R}
\hat{H}_{A+B} = \frac{1}{2}\omega_{a}\hat{\sigma}_{z} + \omega_{b}\hat{b}^{\dagger}\hat{b} + g \hat{\sigma}_{z} (\hat{b}+\hat{b}^{\dagger}),
\end{equation}
$\omega_{a}$ ($\omega_{b}$) being the frequency of the two-level system  (resonator), and $g$ is the coupling strength between the two-level system (TLS) and resonator (R). In this case, $\hat{a}^{(\dagger)}$ is replaced by the respective Pauli operators $\hat{\sigma}_{-(+)}$ in Eq.~(\ref{eq:sys-bath}). This longitudinal spin-boson interaction between a qubit and a resonator can be realized in the circuit QED~\cite{PhysRevLett.115.203601, PhysRevB.93.134501} or in the electro-mechanical systems~\cite{LaHaye2009, PhysRevA.93.063861}. In the case of circuit QED, subsystem B is a single electromagnetic mode of a microwave resonator, and in the electro-mechanical system, it is a mode of nanomechanical resonator. Accordingly, subsystem B can be a photon or phonon mode depending on the choice of the system.\\
(ii) If A and B are both resonators and interact via optomechanical coupling, then
\begin{equation}
\hat{H}_{A+B} = \omega_{a}\hat{a}^\dagger\hat{a} + \omega_{b}\hat{b}^{\dagger}\hat{b} - g \hat{a}^\dagger\hat{a} (\hat{b}+\hat{b}^{\dagger}),
\end{equation} 
 Typically, subsystem B is a micromechanical resonator in the optomechanical systems~\cite{RevModPhys.86.1391}, accordingly B represents a phonon mode. 
  However, the optomechanical-like coupling has been theoretically proposed~\cite{PhysRevA.90.053833} and experimentally realized in the circuit QED~\cite{PhysRevLett.120.227702,Bothner2021}, where both A and B represent the photon modes. Optomechanical-like coupling can be realized in various setups including microtoroids~\cite{Schliesser2008}, levitated particles~\cite{Millen_2020}, and cavity magnomechanical system~\cite{Zhange1501286}. 

We note that our scheme is valid in general for a system in which A and B modes interact dispersively through the Hamiltonian $\hat{H}_{A+B}=g\hat{N}_{0}\hat{X}$, where $\hat{N}_{0}=\zeta\hat{H}_{A}$ with $\zeta$ being a positive constant, and $\hat{X}$ is observable of mode B.
\section{Model Analysis}\label{sec:MA}
In this section, we shall analyze the model proposed in the previous section by deriving the master equation and associated Fokker-Plank equation of the reduced subsystem B. In our analysis, we consider parameters based on the circuit QED realization of the isolated system $\hat{H}_{A+B}$~\cite{LaHaye2009,PhysRevLett.115.203601} unless otherwise stated: $\omega_{a}=2\pi\times 10$ GHz, $\omega_{b}=2\pi\times 500$ MHz, $\kappa_{h}=\kappa_{c}=2\pi\times 200$ MHz, $\kappa_{b}=2\pi\times 1$ KHz, and $g=2\pi\times 20$ MHz. {\color{blue} The numerical results are obtained using Python open-source package QuTip \cite{JOHANSSON20131234}.}
In a recent experimental work, it has been demonstrated that the single-photon coupling $g$ can be reached to $10\%$ of the maximum decay rate in the system~\cite{Bothner2021}. Hence, a strong coupling regime can be realized within the state-of-the-art experimental setups.
In what follows, we consider the TLS-R system for the illustration of our scheme, similar results can also be obtained for R-R interaction.  
\subsection{The master equation}\label{sec:ME}
To derive the master equation, we first diagonalize the isolated system Hamiltonian $\hat{H}_{A+B}$ using the transformation~\cite{Lang1963, PhysRevE.90.022102, Naseem_2020}
\begin{equation}
\hat{U} = e^{-\eta\hat{\sigma}_{z}(\hat{b}^{\dagger}-\hat{b})},
\end{equation}
here $\eta=g/\omega_{b}$. The diagonalized Hamiltonian shows mode A and B frequencies are unaffected:
\begin{equation}
\tilde{H} = \omega_{a}\tilde{\sigma}_{+}\tilde{\sigma}_{-} + \omega_{b}\tilde{b}^{\dagger}\tilde{b}-\frac{g^2}{\omega_{b}}.
\end{equation}
The transformed operators are given by
\begin{align}
\tilde{\sigma}_{-} &= \hat{\sigma}_{-}e^{-\eta(\hat{b}^{\dagger}-\hat{b})}, \\
\tilde{b} &= \hat{b} - \eta\hat{\sigma}_{z}.
\end{align}
The master equation can be derived by transforming these operators into the interaction picture followed by the standard Born-Markov and secular approximations. In addition, by ignoring all higher-order terms $\mathcal{O}(\eta^4)$, the resulting master equation is given by~\cite{PhysRevE.90.022102, PhysRevA.98.052123, Naseem_2020}
\begin{figure}[t!]
  \centering
  \includegraphics[scale=0.52]{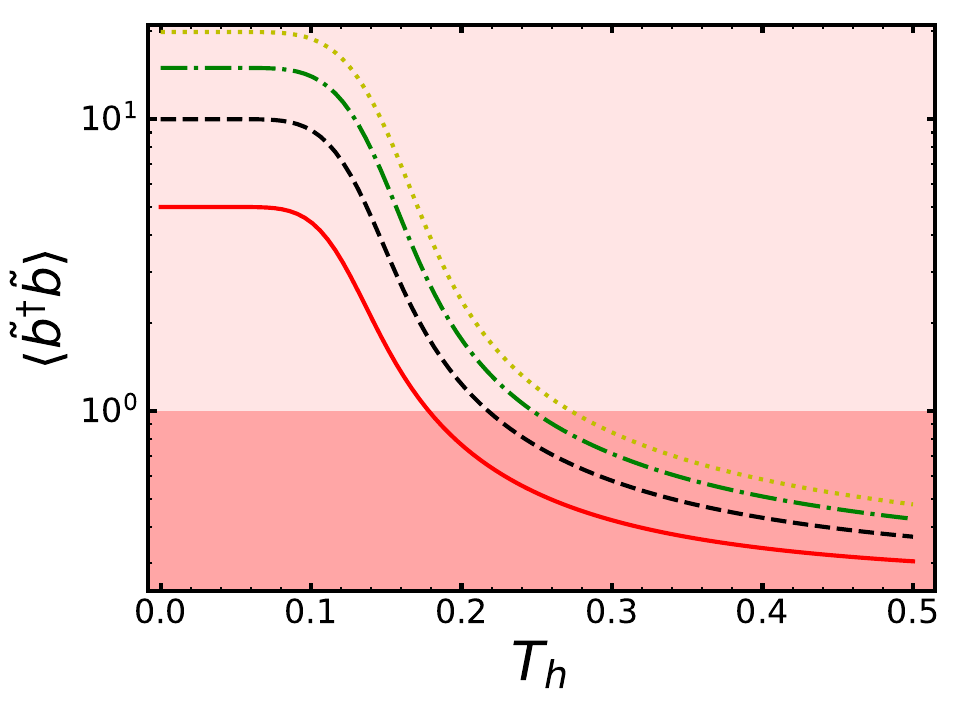}
  \caption{Two-photon cooling. Stationary mean phonon number $\langle \tilde{b}^{\dagger}\tilde{b}\rangle$ of the resonator as a function of the hot bath temperature $T_{h}$ {\color{blue}obtained by the numerical solution of the full master equation~(\ref{eq:filtMEM})} for different thermal occupation number $\bar{n}_{b}$: $\bar{n}_{b}= 5$ (solid red line), $\bar{n}_{b}= 10$ (dashed black line), $\bar{n}_{b}= 15$ (dot-dashed green line), and $\bar{n}_{b}= 20$ (dotted yellow line). The other parameters are: $\omega_{a}=1$, $\omega_{b}=0.05$, $g = 0.005$, $\kappa_{h}=0.02$, $\kappa_{c}=0.02$, $\kappa_{b}=0.0002$, and $T_{c}=0$. All the parameters are scaled with the TLS frequency $\omega_{a}=2\pi\times10$ GHz.}
  \label{fig:fig2}
\end{figure}
\begin{equation}\label{eq:ME}
\frac{d\tilde{\rho}}{dt} = \tilde{\mathcal{L}}_{\alpha=H,C} + \tilde{\mathcal{L}}_{R},
\end{equation}
here $\tilde{\mathcal{L}}_{\alpha=H,C}$ and $\tilde{\mathcal{L}}_{R}$ are the Liouville super-operators for the hot, cold and resonator B baths, respectively, and given by
\begin{align}
	\label{eq:L_L}
\tilde{ \mathcal{L}}_{\alpha} &= G_{\alpha}(\omega_{a})\{D[\tilde{\sigma}_{-}]+\eta^{3}D[\tilde{\sigma}_{-}\tilde{b}^{\dagger}\tilde{b}]\}
	\\ \nonumber & + G_{\alpha}(-\omega_{a})\{D[\tilde{\sigma}_{+}] + +\eta^{3}D[\tilde{\sigma}_{+}\tilde{b}^{\dagger}\tilde{b}]\}
	 \\ \nonumber &+ \sum_{n=1,2}\eta^{n+1}\bigg\{G_{\alpha}(\omega_{a}-n\omega_{b})D[\tilde{\sigma}_{-}\tilde{b}^{\dagger n}]
	\\ \nonumber & + G_{\alpha}(-\omega_{a}+n\omega_{b})D[\tilde{\sigma}_{+}\tilde{b}^{n}]
	\\ \nonumber &+  G_{\alpha}(\omega_{a}+n\omega_{b})D[\tilde{\sigma}_{-}\tilde{b}^{n}]
	\\ \nonumber &+ G_{\alpha}(-\omega_{a}-n\omega_{b})D[\tilde{\sigma}_{+}\tilde{b}^{\dagger n}]\bigg\},
	\\ 
	\tilde{\mathcal{L}}_{R}&= G_{R}(\omega_{b})D[\tilde{b}]
	+ G_{R}(-\omega_{b})D[\tilde{b}^{\dagger}]. \label{eq:L_M}
\end{align} 
Here $D[\tilde{o}]$ is the Lindblad dissipator defined as
\begin{equation}
D[\tilde{o}] = \frac{1}{2}(2\tilde{o}\tilde{\rho}\tilde{o}^{\dagger}-\tilde{\rho}\tilde{o}^{\dagger}\tilde{o}-\tilde{o}^{\dagger}\tilde{o}\tilde{\rho}),
\end{equation}
and $G_{j}(\omega)$ is the bath spectral response function. In this work, we consider one dimensional Ohmic spectral densities of the baths, given by
\begin{eqnarray}\label{eq:SRF}
G_{j}(\omega)=
\begin{cases}
\kappa_{j}(\omega)[1 + \bar{n}_{j}(\omega)] &\omega> 0, \\
\kappa_{j}(\abs{\omega})\bar{n}_{j}(\abs{\omega}) &\omega< 0, 
\end{cases}
\end{eqnarray}
and $G_{j}(0)=0$ for the Ohmic spectral densities of the baths. Here $\kappa_{j}(\omega)$ denotes the system-bath coupling strength. The mean number of quanta in the thermal baths is described by $\bar{n}_{i}(\omega)= 1/(e^{\omega/T_{j}}-1)$ ( We take Boltzmann constant $k_{B}=1$).

Recall that in the case of TLS-R coupling, subsystem B can be a photon mode of a microwave resonator~\cite{PhysRevLett.115.203601} or a phonon mode of a micromechanical resonator~\cite{LaHaye2009,PhysRevA.93.063861}. In what follows, we refer to B as a photon mode for convenience. 
It has been shown previously~\cite{LaHaye2009, PhysRevB.93.134501} that if the modes A and B interact via longitudinal coupling, i.e., $g \hat{\sigma}_{z}(\hat{b}+\hat{b}^{\dagger})$, a coherent external drive of frequency $\omega_{L}=\abs{\omega_{a}\pm n\omega_{b}}$ on mode A induces sideband transitions of the order $n$. This can be employed for the cooling of resonator B. For instance, a coherent drive at the first lower sideband ($\omega_{L}=\omega_{a}-\omega_{b}$) may lead to the preparation of vibrational ground-state of B via dynamical backaction sideband cooling~\cite{PhysRevLett.99.093901, PhysRevLett.99.093902, PhysRevA.77.033804,Schliesser2008, Teufel2011}. 
Similarly, a coherent drive at the lower second-order sideband ($\omega_{L}=\omega_{-}=\omega_{a}-2\omega_{b}$) leads to two-photon (phonon) cooling of the resonator B~\cite{PhysRevA.82.021806, PhysRevA.95.053844}. The coherent drive at $\omega_{-}$ results in an effective parametric amplifier Hamiltonian $g(\hat{\sigma}_{-}\hat{b}^{\dagger 2} + \hat{\sigma}_{+}\hat{b}^2)$. It is known to destroy two quanta in the mode B by adding an excitation in the pump mode A, which leads to two-photon cooling process~\cite{PhysRevA.82.021806}.
 In addition to cooling, such Hamiltonians can be exploited to change the statistical properties of the mode B from super-Poissonian to sub-Poissonian~\cite{PhysRevA.82.021806, PhysRevA.95.053844}, generation of a macroscopic superposition state~\cite{PhysRevA.88.023817, Asjad_2014}, and realization of the photon or phonon blockade~\cite{PhysRevA.92.023838,  PhysRevA.93.063861, PhysRevB.87.235319}. In all these proposals,  the two-photon process needs to dominate one-photon processes~\cite{Dodonov_1997, Dodonov_1997_2}.

In our scheme, there is no coherent source, instead, the TLS is driven by two incoherent drives. Due to the presence of these thermal drives and longitudinal interaction between TLS-R, both upper and lower photon sidemodes are present in the master equation~(\ref{eq:L_L}). By analogy with the coherent two-photon process, we may want to drive the TLS at the second lower sideband and suppress all other sidebands in Eq.~(\ref{eq:L_L}). Consequently, the two-photon absorption (cooling) process dominates the two-photon emission (heating), one photon absorption, and emission processes. This can be achieved using a quantum reservoir engineering method namely bath spectrum filtering~\cite{doi:10.1080/09500349414550381, PhysRevLett.109.090601, PhysRevE.87.012140, PhysRevE.90.022102, Ghosh12156, PhysRevResearch.2.033285, Naseem_2020, Naseem2021}.  To this end, we consider the filtered bath spectra of the hot and cold bath shown in Fig.~\ref{fig:fig1}(b). The hot thermal bath plays the role of a coherent drive. It couples only to a transition frequency of $\omega_{-}$ and coupling to all other transition frequencies is negligible due to bath filtering. On contrary to the coherent two-photon cooling process, in which the driving mode has a single incoherent environment, we require an additional thermal bath coupled to TLS [see Fig.~\ref{fig:fig1}]. The need for the additional cold bath is imposed by the second law of thermodynamics~\cite{Naseem2021}. This bath can only induce the transition of frequency $\pm \omega_{a}$.  

If we consider baths spectra shown in Fig.~\ref{fig:fig1}, the master equation (\ref{eq:L_L}) reduces to
\begin{eqnarray}
\tilde{\mathcal{L}}_{C} &=& \tilde{G}_{C}(\omega_{a})\{D[\tilde{\sigma}_{-}] + \eta^{3}D[\tilde{\sigma}_{-}\tilde{b}^{\dagger}\tilde{b}]\}\nonumber \\ 	
	&+& \tilde{G}_{C}(-\omega_{a})\{D[\tilde{\sigma}_{+}] + \eta^{3}D[\tilde{\sigma}_{+}\tilde{b}^{\dagger}\tilde{b}]\},\nonumber \\ 	
\tilde{\mathcal{L}}_{H}	&=& \eta^3\big(\tilde{G}_{H}(\omega_{-})D[\tilde{\sigma}_{-}\tilde{b}^{\dagger 2}]
	+ \tilde{G}_{H}(-\omega_{-})D[\tilde{\sigma}_{+}\tilde{b}^2]\big),\nonumber \\ 
	\tilde{\mathcal{L}}_{R}&=& G_{R}(\omega_{b})\tilde{\mathcal{D}}[\tilde{b}]
	+ G_{R}(-\omega_{b})\tilde{\mathcal{D}}[\tilde{b}^{\dagger}].\label{eq:filtMEM}
\end{eqnarray}
\begin{figure}[t!]
  \centering
  \includegraphics[scale=0.52]{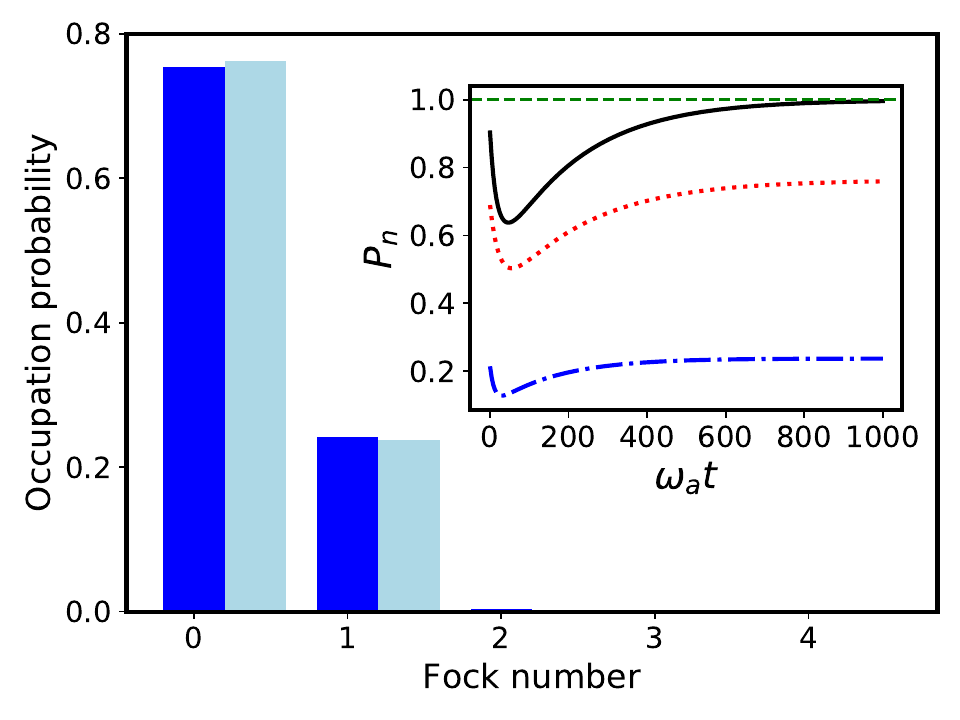}
  \caption{Photon-number distribution $P_{n}$ for high temperatures. Dark and light blue bars represent photon statistics obtained by the numerical solution of the full master equation~(\ref{eq:filtMEM}), and analytical solution [Eq.~(\ref{eq:populations})] evaluated from Fokker-Plank equation ~(\ref{eq:FP}), respectively. The inset shows probabilities of measuring $n$ photons as a function of scaled time $\omega_{a}t$. Dotted red, dot-dashed blue, and solid black lines are for $P_{0}$, $P_{1}$, and $P_{0}+P_{1}$, respectively. The dashed green line shows total sum $\sum_{n} P_{n}=1$.  
   The other parameters are: $\omega_{a}=1$, $\omega_{b}=0.05$, $g = 0.005 $, $\kappa_{h}=0.02$, $\kappa_{c}=0.02$, $\kappa_{b}=0.0002$, $\bar{n}_{b}=5$, $T_{c}=0$, and $T_{h}=2$. All the parameters are scaled with the TLS frequency $\omega_{a}=2\pi\times10$ GHz.}
  \label{fig:fig3}
\end{figure}
The bath spectral density of the resonator heat bath is not filtered, accordingly, the Liouville super-operator $\tilde{\mathcal{L}}_{R}$ remains unchanged. The filtered bath spectrum $\tilde{G}_{\alpha}(\omega)$ is given by~\cite{doi:10.1080/09500349414550381, PhysRevE.90.022102, PhysRevResearch.2.033285}
\begin{equation}
\tilde{G}_{\alpha} = \frac{\kappa^{f}_{\alpha}}{\pi}\frac{(\pi G_{\alpha}(\omega))^2}{(\omega-(\omega^{f}_{\alpha}+\Delta^{L}_{\alpha}(\omega)))^2+(\pi G_{\alpha}(\omega))^2},
\end{equation}   
$\kappa^{f}_{\alpha}$ being the coupling rate of the TLS with the filter, and $\omega^{f}_{\alpha}$ is the bath spectrum resonance frequency. The modes closer to the resonance frequency are more strongly coupled to the system. The bath-induced Lamb shift is given by
\begin{equation}
\Delta^{L}_{\alpha}(\omega) = P \int^{\infty}_{0} d\omega^{'} \frac{G_{\alpha}(\omega^{'})}{\omega-\omega^{'}},
\end{equation}
and P being the principal value.
\subsection{Focker-Plank equation}\label{sec:FP}
It can be advantageous to map the quantum master equation (\ref{eq:filtMEM}) into a classical stochastic process with appropriate phase space representation. This can be done by deriving a Focker-Plank equation from Eq.~(\ref{eq:filtMEM}).
We are interested in the dynamics of the resonator, therefore upon taking trace over the TLS, the reduced master equation for the resonator takes the form
\begin{align}\label{eq:RFPE}
\frac{d\tilde{\rho}}{dt} &= \Gamma_{\downarrow} D[\tilde{b}^2] + \Gamma_{\uparrow} D[\tilde{b}^{\dagger 2}] \nonumber \\
& + \gamma_{\downarrow} D[\tilde{b}] + \gamma_{\uparrow} D[\tilde{b}] + \gamma_{d} D[\tilde{b}^{\dagger}\tilde{b}],
\end{align}
here the coupling rates are defined by
\begin{align}
\Gamma_{\downarrow}  &:= \eta^3\tilde{G}_{H}(-\omega_{-})\langle\tilde{\sigma}_{z}+1\rangle, \quad
\Gamma_{\uparrow} := {\eta}^3\tilde{G}_{H}(\omega_{-})\langle\tilde{\sigma}_{z}\rangle,  \nonumber \\
\gamma_{\downarrow} &:= G_{R}(\omega_{b}), \quad \gamma_{\uparrow} := G_{R}(-\omega_{b}), \nonumber \\
\gamma_{d} & := \eta^{3} (\tilde{G}_{C}(\omega_{a})\langle\tilde{\sigma}_{z}\rangle + \tilde{G}_{C}(-\omega_{a})\langle\tilde{\sigma}_{z}+1\rangle).
\end{align}
$\Gamma_{\downarrow}$ ($\gamma_{\downarrow}$) and $\Gamma_{\uparrow}$ ($\gamma_{\uparrow}$) being the two (one)-photon cooling and heating rates, respectively.
The exact analytical solution of Eq.~(\ref{eq:RFPE}) can be found in the limit $\Gamma_{\downarrow}/\Gamma_{\uparrow}\ll 1$, and by ignoring the dephasing $\gamma_{d}$~\cite{Dodonov_1997, Dodonov_1997_2}. Under these approximations, Eq.~(\ref{eq:RFPE}) can readily be transformed into a Fokker-Plank equation~\cite{PhysRevA.95.053844}
\begin{align}\label{eq:FP}
\frac{dP}{dt} = -\sum_{i} \frac{\partial}{\partial\zeta_{i}}[F(\zeta)]_{i}P(\zeta) + \frac{1}{2}\sum_{i,j}\frac{\partial}{\partial\zeta_{i}}\frac{\partial}{\partial\zeta_{j}}[H(\zeta)]_{i, j}P(\zeta),
\end{align} 
here $\zeta = (\mu, \mu^{*})$ and
  \begin{align}
    F(\zeta) &= \begin{bmatrix}
          -\frac{1}{2}\kappa_{b}\mu - \Gamma_{\downarrow}\mu^2\mu^{*}\\    
          -\frac{1}{2}\kappa_{b}\mu^* - \Gamma_{\downarrow}\mu\mu^{* 2} 
         \end{bmatrix},
  \end{align}
  
    \begin{align}
    H(\zeta) &= \begin{bmatrix}
          -\Gamma_{\downarrow}\mu^2 \quad\quad    - \kappa_{b}\bar{n}_{b}\\    
          -\kappa_{b}\bar{n}_{b}    \quad\quad    - \Gamma_{\downarrow}\mu^{* 2} 
         \end{bmatrix},
  \end{align}
$F(\zeta)$, and $H(\zeta)$ being the drift vector and diffusion matrix, respectively. In our scheme, in the limit $T_{H}\gg T_{R}> T_{C}$, and $T_{C}\approx 0$,  two-photon amplification becomes negligibly small compared to two-photon cooling, i.e., $\Gamma_{\uparrow}\ll\Gamma_{\downarrow}$. 
Accordingly, Eq.~(\ref{eq:FP}) can be used to analyze the dynamics of the resonator in this regime. A comparison between results obtained from numerical simulation of the full master equation~(\ref{eq:filtMEM}) and analytical results from the Fokker-Plank equation~(\ref{eq:FP}) are presented in the next section. 
\begin{figure}[t!]
  \centering
  \includegraphics[scale=0.52]{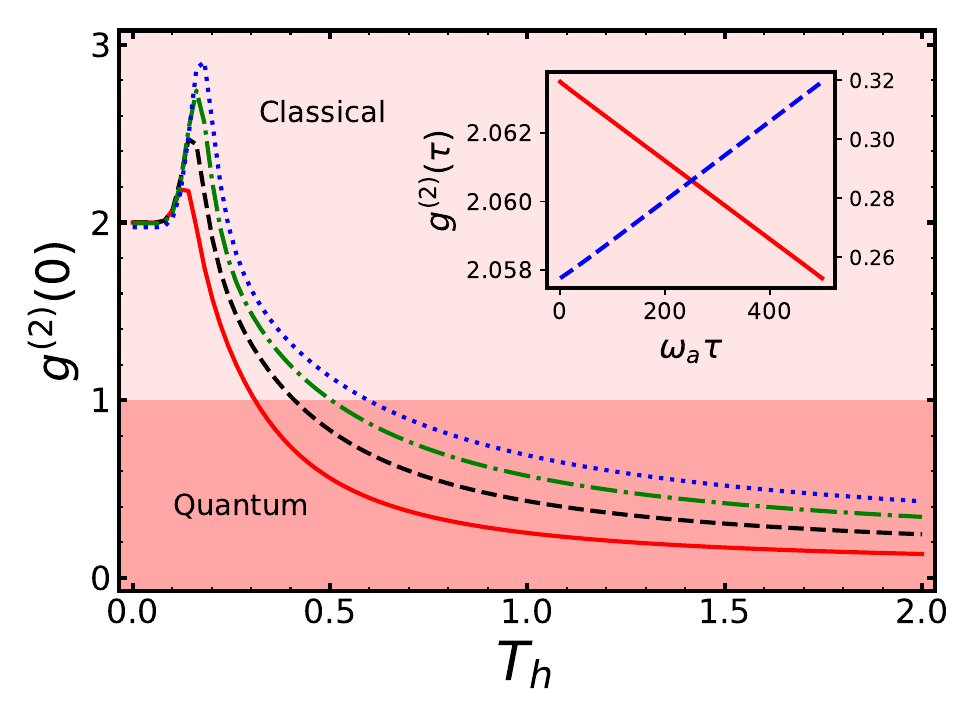}
  \caption{Antibunching and sub-Poissonian statistics. Steady state zero-time-delay second order coherence function $g^{(2)}(0)$ as a function of the hot bath temperature $T_{H}$ for different thermal occupation number $\bar{n}_{b}$: $\bar{n}_{b}= 5$ (solid red line), $\bar{n}_{b}= 10$ (dashed black line), $\bar{n}_{b}= 15$ (dot-dashed green line), and $\bar{n}_{b}= 20$ (dotted blue line). The inset shows the second-order correlation function $g^{(2)}(\tau)$ with a delay time $\tau$ as a function of scaled time $\omega_{a} \tau$ for different values of hot bath temperature $T_{H}$. {\color{blue} The results are obtained by the numerical solution of the full master equation~(\ref{eq:filtMEM}).}
   Red solid line, and blue dashed lines are for $T_{H}=0.1$, and $T_{H}=1$, respectively. 
  The other parameters are: $\omega_{a}=1$, $\omega_{b}=0.05$, $g = 0.01$, $\kappa_{h}=0.02$, $\kappa_{c}=0.02$, $\kappa_{b}=0.0002$, and $T_{c}=0$. All the parameters are scaled with the TLS frequency $\omega_{a}=2\pi\times10$ GHz.}
  \label{fig:fig4}
\end{figure}
\section{Results}\label{sec:results}
\subsection{Two-photon (phonon) cooling}
 If we employ bath spectrum filtering and consider the baths spectra of the hot and cold baths as shown in Fig.~\ref{fig:fig1}(b), the resonator can be cooled by two-photon cooling process. This is only possible if the system parameters are considered in the limit $T_{H}\gg T_{R}> T_{C}$ and $\kappa_{b}\bar{n}_{b}\ll\Gamma_{\downarrow}$. For these system parameters,  two-photon amplification and one photon processes due to resonators' bath become much smaller than the two-photon cooling rate $\Gamma_{\downarrow}$. In the cooling process, two quanta of energy is removed from the resonator and it is dumped into the cold bath. The energy required for this process is provided by the hot bath, and this process results in the cooling of the resonator.  The preceding analysis is confirmed by Fig.~\ref{fig:fig2}, in which the stationary mean photon number of the resonator is plotted as a function of the hot bath temperature $T_{H}$, for different thermal occupation number $\bar{n}_{b}$. The increase in  $T_{H}$ results in the increase in the two-photon cooling rate $\Gamma_{\downarrow}$, that leads to the cooling of the resonator. For high temperatures $T_{H}\gg\{\omega_{a}, \omega_{b}, T_{R}\}$, nonlinear two-photon cooing is the dominant source of damping in the system.
On contrary to one-photon cooling~\cite{Naseem_2020}, ground-state cooling is not possible via two-photon cooling process~\cite{PhysRevA.82.021806}. 

In the limit of strong two-photon cooling rate and weak one photon linear damping, i.e., $\kappa_{b}\bar{n}_{b}\ll\Gamma_{\downarrow}$, only the ground and first excited states of the resonator are significantly populated. It is because of the nature of two-photon process. All the components of the odd-photon-number state collapse to the first excited state and even numbers to the ground state~\cite{PhysRevA.48.1582}. In the limit of strong two-photon cooling rate, the populations of the ground and first excited states can be evaluated from Eq.~(\ref{eq:FP}) and given by~\cite{PhysRevA.82.021806, PhysRevA.95.053844}
\begin{align}\label{eq:populations}
P(0) &\approx \frac{3\bar{n}_{b}+1}{4\bar{n}_{b}+1}, \nonumber \\ 
P(1) &\approx \frac{\bar{n}_{b}}{4\bar{n}_{b}+1}.
\end{align} 
A comparison of numerical results of photon-number distribution $P_{n}$ calculated by solving the full master equation~(\ref{eq:filtMEM}) and analytical results [Eq.~(\ref{eq:populations})] is presented in Fig.~\ref{fig:fig3}. Our numerical results show an good agreement with the analytical results. We find that the sum of probabilities of measuring photon in the ground and first excited state $P_{0}+P_{1}$ is almost one. 

\subsection{Photon (phonon) antibunching}
We now proceed to analyze the photon statistics of the resonator. To this end, we employ second-order correlation function defined for the stationary state
\begin{equation}\label{eq:twotime}
g^{2}(\tau) = \frac{\langle\tilde{b}^{\dagger}(0)\tilde{b}^{\dagger}(\tau)\tilde{b}(\tau)\tilde{b}(0)\rangle}{\langle\tilde{b}^{\dagger}(0)\tilde{b}(0)\rangle^2},
\end{equation}
$\tau$ being the delay time between two measurements. According to this standard definition, photon field is antibunched (bunched) if $g^{2}(0)<g^{2}(\tau)$ [$g^{2}(0)>g^{2}(\tau)$] for positive delay time $\tau >0$. For zero-time delay $\tau= 0$ second-order correlation function provides information on the photon sub-Poissonian and super-Poissonian statistics, and it is defined as
\begin{equation}
g^{2}(0) = \frac{\langle\tilde{b}^{\dagger 2}\tilde{b}^2\rangle}{\langle\tilde{b}^{\dagger}\tilde{b}\rangle^2}.
\end{equation}
We refer to statistics of the photon field being sub-Poissonian (super-Poissonian) if $g^{2}(0)<1$ [$g^{2}(0)>1$], which indicates non-classical (classical) state of the photon field~\cite{MandelBook}. In addition, for a coherent source $g^{2}(0)=1$. We stress that although antibunching and sub-Poissonian photon statistics tend to occur together and reveal certain quantum features of the photon state, these are not one and the same~\cite{PhysRevA.41.475}. The two-time photon correlations define bunching and antibunching, while super-Poissonian and sub-Poissonian statistics are given by single-time photon correlations.

In Fig.~\ref{fig:fig4}, we plot $g^{2}(0)$ as a function of the hot bath temperature $T_{H}$ for different thermal occupation number $\bar{n}_{b}$. The results are obtained by solving the the full master equation~(\ref{eq:filtMEM}). This figure confirms that starting from a super-Poissonian statistics, photon field of the resonator attains a sub-Poissonian statistics if driven by a heat bath at higher temperatures. This indicates that sub-Poissonian statistics of the photon field can be obtained in our scheme in the limit $T_{H}\gg T_{R}>T_{C}$. We note that if thermal damping of the resonator  is large $\kappa_{b}\bar{n}_{b}>\Gamma_{\downarrow}$, then second-order coherence function $g^{2}(0)$ is always greater than one, which indicates the resonator is in a thermal state~\cite{PhysRevA.95.053844}. The second-order correlations function with a delay time $\tau$ is plotted in the inset of Fig.~\ref{fig:fig4} as a function of scaled time $\omega_{a} t$. It shows that, for low hot bath bath temperature $T_{H}=0.1$,  $g^{2}(0)>g^{2}(\tau)$ (red solid line) that indicates the absence of antibunching in the state of the resonator. For higher temperature $T_{H}=1$, $g^{2}(0)<g^{2}(\tau)$ (blue dashed line) which reveals the quantum features of the resonator state. 

Our results are in agreement with the previously reported works, which show that by using parametric amplifier interaction $g(\hat{a}\hat{b}^{\dagger 2}+\hat{a}^{\dagger}\hat{b}^2)$ between the subsystems A and B, the mode B can attain sub-Poissonian statistics provided a sufficiently strong coupling rate between the two modes~\cite{PhysRevA.95.053844}. Alternatively, dissipative two-photon absorption process has been shown to change the statistics of the photon field from super-Poissonian to sub-Poissonian~\cite{RevModPhys.54.1061}. 

\section{Conclusions}\label{sec:conc}
We have investigated statistics of the photon (phonon) field under linear and nonlinear dampings in a coupled bipartite system attached to three independent heat baths. It has shown that a nonlinear two-photon (phonon) cooling by heating process can be realized from the linear system-bath couplings. This is achieved by employing bath spectrum filtering. For high temperatures of these heat baths, strong photon (phonon) antibunching and sub-Poissonian statistics are reported. Our key result is that the antibunching in the photon (phonon) field increased with the increase in spectrally filtered heat bath temperature. 
This notion of emergence and increase in quantumness by environment temperature is applied to a two-level system coupled longitudinally with a harmonic oscillator or analogous optomechanical system. The underlying physical mechanism is explained by showing that the two-photon cooling process is the dominant source of damping. Our analysis may provide a possible route for the realization of the thermally controlled nonlinear damping in the quantum systems, and generation of profound quantum states via cooling by heating schemes. {\color{blue} The thermal environment may destroy the quantumness of the quantum states of photon/phonon fields. Consequently, our results can be of fundamental and practical interest for presenting a route to high-temperature quantumness and single-photon/phonon sources \cite{doi:10.1063/1.1650032, O'Brien2009}.} 

\bibliography{antibunch}

\end{document}